# MINIMIZATION OF HANDOFF LATENCY BY CO-ORDINATE EVALUATION METHOD USING GPS BASED MAP

Debabrata Sarddar[1], Joydeep Banerjee[1], Souvik Kumar Saha[1], Tapas Jana[2], Utpal Biswas[3], M.K. Naskar[1]

1. Department of Electronics and Telecommunication Engg, Jadavpur University, Kolkata – 700032. E-mail:dsarddar@rediffmail.com, jogs.1989@rediffmail.com, souviksaha@ymail.com, mrinalnaskar@yahoo.co.in .

2. Department of Electronics and Communication Engg, Netaji Subhash Engg College, Techno City, Garia, Kolkata – 700152. Email: tjanansec@gmail.com,

3. Department of Computer Science and Engg, University of Kalyani, Nadia, West Bengal, Pin-741235, Email: utpal01in@yahoo.com

***ABSTRACT***

*Handoff has become an essential criterion in mobile communication system, specially in urban areas, owing to the limited coverage area of Access Points (AP). Handover of calls between two BS is encountered frequently and it is essentially required to minimize the delay of the process. Many solutions attempting to improve this process have been proposed but only a few use geo-location systems in the management of the handover. Here we propose to minimize the handoff latency by minimizing the number of APs scanned by the mobile node (MN) during each handoff procedure. We consider the whole topographical area as a two dimensional plane. By GPS, we can note down the co-ordinates of the MN at any instant. The average rate of change of its latitudinal distance and longitudinal distance with a specific time period is evaluated at the end of the given time period. With the knowledge of the given parameter, it is possible to determine the latitude and longitude of the MN after a particular instant of time. Hence the direction of motion of the MN can be determined which in turns gives the AP towards which the MN is heading towards. This reduces the number of APs to be scanned. Thus, on an overall basis, the handoff latency can be reduced by almost half to one third of its value.*

## 1. INTRODUCTION

In recent years, IEEE 802.11 based wireless local area networks (WLAN) have been widely deployed for business and personal applications. The main issue regarding wireless network technology is handoff management. Quality of service (QoS) demanding applications like Voice over IP (VoIP) and multimedia need seamless handover. Many techniques have been proposed to improve the link layer 2 handover.

IEEE 802.11b based wireless and mobile networks [1], also called Wi-Fi commercially, are experiencing a very fast growth upsurge and are being widely deployed for providing variety of services as it is cheap, and allows anytime, anywhere access to network data. However they suffer from limited coverage range of AP, resulting in frequent handoffs, even in moderate mobility scenarios.

With the advent of real time applications, the latency and packet loss caused by mobility became an important issue in Mobile Networks. The most relevant topic of discussion is to reduce the IEEE 802.11 link-layer handoff latency. IEEE 802.11 MAC specification [3] defines two operation modes: ad hoc and infrastructure mode. In the ad hoc mode, two or more stations (STAs) recognize each other through beacons and hence establish a peer-to-peer relationship. In infrastructure mode, an AP provides network connectivity to its associated STAs to form a Basic Service Set (BSS). Multiple APs form an Extended Service Set (ESS) that constructs the same wireless networks. We now describe the handoff procedure with its various phases.

## 1.1 HANDOVER PROCESS

The complete handoff procedure can be divided into 3 distinct logical parts: scanning, authentication and re-association. In the first phase, an STA scans for AP's by either sending Probe Request messages or by listening for beacon message. After scanning all channels, an AP is selected using the Received Signal Strength Indication (RSSI) and CI ratio, and the selected AP exchanges IEEE 802.11 authentication messages with the STA. Finally, if the AP authenticates the STA, the STA sends Re-association Request message to the new AP.

### 1.1.1. SCANNING

Scanning can be divided into active and passive scans. During an active scan, the STA broadcasts a probe request packet asking all APs in those specific channels to impart their existence and capability with a probe response package. In a passive scan, the STA listens passively for the beacons bearing all necessary informations like beacon interval, capability information, supported rate etc. about an AP. Active scan is normally speedy as it aims to bypass the most time consuming phases in the layer (L2) handoff procedure, but is unreliable, since probe packets may get lost or greatly delayed in wireless traffic jams. Passive scan, though reliable, has a long waiting time for beacons which is prohibitive to many services. Thus a selective channel probing should be judiciously used. The active scans introduce two parameters:

'Min Channel Time' represents the arrival time of the first probe response. So a client must listen for this period of time to decide whether there are any APs on this channel. It is recommended to be set as 3-7 ms. 'Max Channel Time' is the estimated time to collect all probe responses. It is supposed to be of the magnitude of tens of milliseconds. For all practical implementation, the maximum channel time is set to 30 ms [7].

### 1.1.2. AUTHENTICATION

Authentication is necessary prior to association. Authentication must either immediately proceed to association or must immediately follow a channel scan cycle. In pre-authentication schemes, the MN authenticates with the new AP immediately after the scan cycle finishes. IEEE 802.11 defines two subtypes of authentication service: 'Open System' which is a null authentication algorithm and 'Shared Key' which is a four-way authentication mechanism. If Inter Access Point Protocol (IAPP) is used, only null authentication frames need to be exchanged in the re-authentication phase. Exchanging null authentication frames takes about 1-2ms.

### 1.1.3. RE-ASSOCIATION

Re-association is a process for transferring associations from one AP to another. Once the STA has been authenticated with the new AP, re-association can be started. Previous works has shown re-association delay to be around 1-2 ms. The overall delay is the summation of scanning delay, authentication delay, and re-association delay.

According to [7], 90% of handoff delay comes from scanning delay. The range of scanning delay is given by:-

N × Tmin _ Tscan _ N × Tmax

where N is the total number of channels according to the spectrum released by a country, Tmin is Min Channel Time, Tscan is the total measured scanning delay, and Tmax is Max Channel Time. Here we focus on reducing the scanning delay by minimizing the total number of scans performed. The total handoff process is shown is Figure 1.

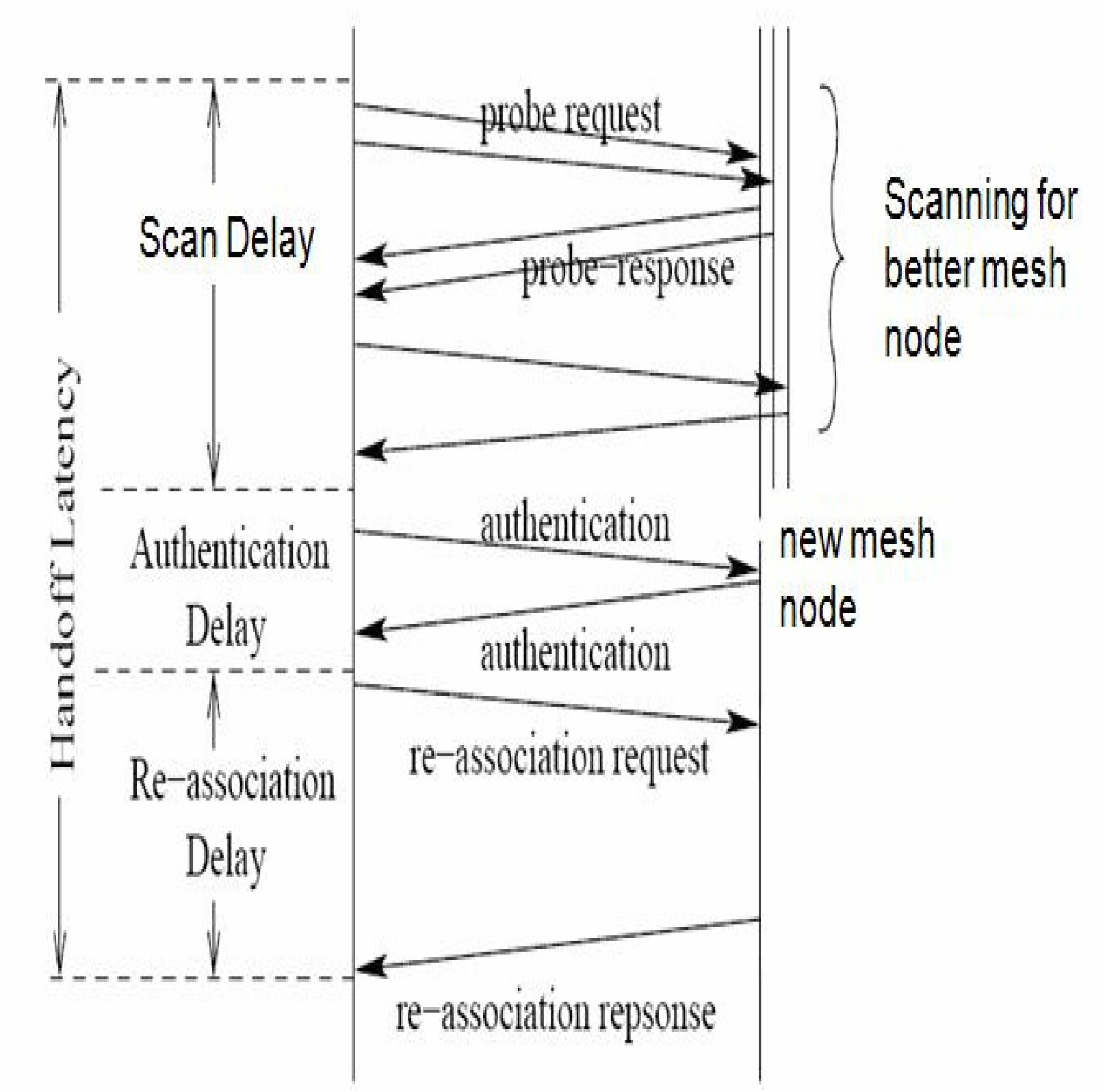

Figure 1.The total handoff process in brief.

## 1.2. GLOBAL POSITIONING SYSTEM

**Global Positioning System** (**GPS**) is a space-based global navigation satellite system. It provides reliable positioning, navigation, and timing services to worldwide users on a continuous basis in all weather, day and night, anywhere on or near the Earth which has an unobstructed view of four or more GPS satellites.GPS is made up of three segments: Space, Control and User. The Space Segment is composed of 24 to 32 satellites in Medium Earth Orbit and also includes the boosters required to launch them into orbit. GPS satellites broadcast signals from space that GPS receivers use to provide three-dimensional location (latitude, longitude, and altitude) plus precise time. GPS has become a widely used aid to navigation worldwide, and is a useful tool for map-making, land surveying, commerce, scientific uses, tracking and surveillance, and hobbies such as geocaching and way marking.

A GPS receiver calculates its position by precisely timing the signals sent by the GPS satellites high above the Earth. Each satellite continually transmits messages which include

- the time the message was transmitted
- precise orbital information (the ephemeris)
- the general system health and rough orbits of all GPS satellites (the almanac).

The receiver utilizes the messages it receives to determine the transit time of each message and computes the distances to each satellite. These distances along with the satellites' locations are used with the possible aid of dilatation to compute the position of the receiver. This position is then displayed, perhaps with a moving map display or latitude and longitude; elevation information may be included. Many GPS units also show derived information such as direction and speed, calculated from position changes. Using messages received from a minimum of four visible satellites (this is because four spheres intersect in space at a minimum of one point, thus specifying a definite location), a GPS receiver is able to determine the times sent and then the satellite positions corresponding to these times sent. The x, y, and z components of position, and the time sent, are designated as *xi,yi=i,ti* where the subscript *i* is the satellite number and has the value 1, 2, 3, or 4. Knowing the indicated time the message was received *tr*, the GPS receiver can compute the transit time of the message as (*tr-ti*). Assuming the message traveled at the speed of light, c, the distance traveled, *Pi* can be computed as *(tr-ti)c.x*

**1.2.1. SIGNAL ARRIVAL TIME MEASUREMENT**

The position calculated by a GPS receiver requires the current time, the position of the satellite and the measured delay of the received signal. The position accuracy is primarily dependent on the satellite position and signal delay. To measure the delay, the receiver compares the bit sequence received from the satellite with an internally generated version. By comparing the rising and trailing edges of the bit transitions, modern electronics can measure signal offset to within about one percent of a bit pulse width, $\frac{0.01}{(1.023 \times 10^6/sec)}$, or approximately 10 nanoseconds for the C/A code. Since GPS signals propagate at the speed of light, this represents an error of about 3 meters.

This component of position accuracy can be improved by a factor of 10 using the higher-chip rate P(Y) signal. Assuming the same one percent of bit pulse width accuracy, the high-frequency P(Y) signal results in an accuracy of $\frac{(0.01 \times 300,000,000\ m/sec)}{(10.23 \times 10^6/sec)}$ or about 30 centimeters. For the generation of a GPS map, a monitor node travels in a given network and generates an AP map. As monitor node with GPS travels, it receives packet transmitted by the surrounding APs. When certain conditions are met (which will be stated later), the monitor node records the coordinate of the access point with its BSSID (basic service set ID) and the used channel number of the AP. Based on this information, a table is generated consisting of a range of latitudes and longitudes and the used channel number of an AP. When an MN enters the network, it downloads the GPS map from the server. If the distance between the current AP and the monitor node is above a certain threshold, then it sends its current co-ordinate to the server. By this the server finds the minimum distance of the MN from all other APs. Certain APs with distance less than certain threshold are chosen as neighbour. Thus the total number of channels to be scanned is reduced by a great deal. The parameters which consists of AP's coordinates (Latitude and longitude), the IEEE 802.11 channel on which the AP is operating, the APs Service Set Identifier (SSID) and the IPv6 prefix are statically configured. At each position check (performed every second when using the GPS geo-location system), an MN records its current coordinates and compares them to the previous ones in order to determine if it has moved. The distance between two points is calculated using the Haversine formula. It assumes a spherical Earth and ignores ellipsoidal effects but remains particularly well-conditioned for numerical computation even at small distances. Let us denote the previous and the current coordinates of an MN as (lat1, long1) and (lat2, long2) respectively. Let us also denote the latitude separation with _lat and the longitude separation with _long, where angles are in radians, and R the Earth's radius (R = 6, 371km). The distance d between the two points is calculated by the formula:

$$\text{haversin}(d/R) = \text{haversin}(\Delta_{lat}) + \cos(lat1) \times \cos(lat2) \times \text{haversin}(\Delta_{long})$$

where the Haversine function is given by:

$$\text{haversin}(\delta) = \sin^2(\delta/2)$$

Let h denote the haversin(d/R). One can then solve for d either by simply applying the inverse Haversine or by using the arcsin (inverse of sine) function:

$$d = R\times \text{haversin}^{-1}(h) = 2R \times \arcsin(\sqrt{h})$$

If d is greater than 1 meter, we consider that the MN has moved and has to send to the GPS Server a LU message, which includes the identity of its current AP and its current co-ordinates. The MN can know its position information by using a GPS. Building GPS in the MH means the MH is able to track its location continuously within 1 to 2 metre precision. With prediction, it is possible to reduce latency and packet loss. Thus, GPS allows us to anticipate movement calculation with the help of which the need to wait for beacon signals from other FAs is eliminated. Also, handoff target areas can be discovered in advance.

## 2. RELATED WORKS

A lot of researches have been dedicated to improve the handoff performance in IEEE 802.11 based networks. They proposed new algorithms or new protocols. For QoS demanding applications like VoIP and multimedia, seamless handover in mobility support has become a great issue.

In the past few years, some proposed neighbor graph method & some proposed geo-location based handoff procedure. In [2], authors propose selective channel scanning mechanism using neighbor graphs. They scan neighbor APs and collect their respective channel information. They need changes in network infrastructure and use of IAPP. Chung-Sheng Li et all in [3] focus on neighbor graph cache mechanism for link layer2 handover.

These scanning processes involve all the APs in the vicinity of the MN according to some mobility profile, regardless of the MN's motion. As a consequence, a number of APs per MN, e.g. those opposite to the direction of movement, are involved. Therefore, these processes are more power as well as time consuming.

In [4] & [5], authors use GPS based access point maps for handoff management. In [6], S. Kyriazakos et all propose an algorithm to resolve the well known ping-pong and far-away cell effects using the MN's movement and its velocity. They give a brief description of the mechanism without providing any performance evaluation.

J Pesola et all in [7] present a location assisted algorithm to manage handover between WLAN and GPRS networks.

Handoff using received signal strength (RSS) of BS has been proposed previously. Using dynamic threshold value of RSS for handoff management for MTs of different velocities has been described in [10].

# 3. PROPOSED METHOD

Here, we propose to reduce the handover latency by reducing the number of APs scanned by the MN during the handover process. We utilize Global Positioning System (GPS) to implement our mechanism. The selection of the most potential AP by the MN effectively reduces the scanning delay, as the number of channels scanned will be lower.

We assume hexagonal coverage area of an AP with the AP situated at the center. Before introducing the mechanism, we need to describe certain parameters and the block diagram of the environment which we use for our simulation. The mechanism is started after a period of initialization. The GPS response time as been discussed is low as 10 ns with an error of at the most 30 cm. The distance between two APs in urban area is of the range 200 m to 500m. So the initialization phase must be as low as possible. We set the initialization phase to be a maximum of 60 ms. At every 5 ms it records the distance travelled by the MN and adds it with the distance traveled in the previous 5 ms period. So there will be a summation of 12 distances. It can be considered that we can get the distance roughly by this method with an error of approximate 30 cm owing to the use of GPS server. The most important parameter is the average speed of the MN, denoted by the symbol '$s_{avg}$'. It can hence be determined by the sum of the distances calculated in period of 5ms divided by the total time of evaluation i.e. 60 ms. If $\Delta x_i$ is the distance traveled in the $i^{th}$ instance then the speed $s_{avg}$ is given by:

$$s_{avg} = (\sum \Delta x_i)/ 60$$

where 'i' ranges from 1 to 12. This is the initial speed which is required for the further evaluation of other parameters. Note that this average speed changes from time to time as the $s_{avg}$ is calculated in a similar way after the initialization phase. So $s_{avg}$ is a variable quantity. Now, this has an error of approximately 0.5 meters per second to the maximum, because of the distance error of the GPS method. This is very nominal for fast moving vehicles.

Let the approximate maximum handoff delay in the scanning, authentication and re-association phase of a single AP be $t_{delay}$ ms as per the latest proposed algorithm. The potential AP searches can me made up to a certain distance after which the MN performs the handover process. The distance'd' which is required for the rest of handover procedure is given in the Figure 2. So that distance'd' is given by the equation:

$$d = t_{delay} * s_{avg}$$

So,' d' is also a variable parameter. But if at any instant, the MN reaches within this region, even if it varies, the algorithm is stopped and the rest of the procedure starts.

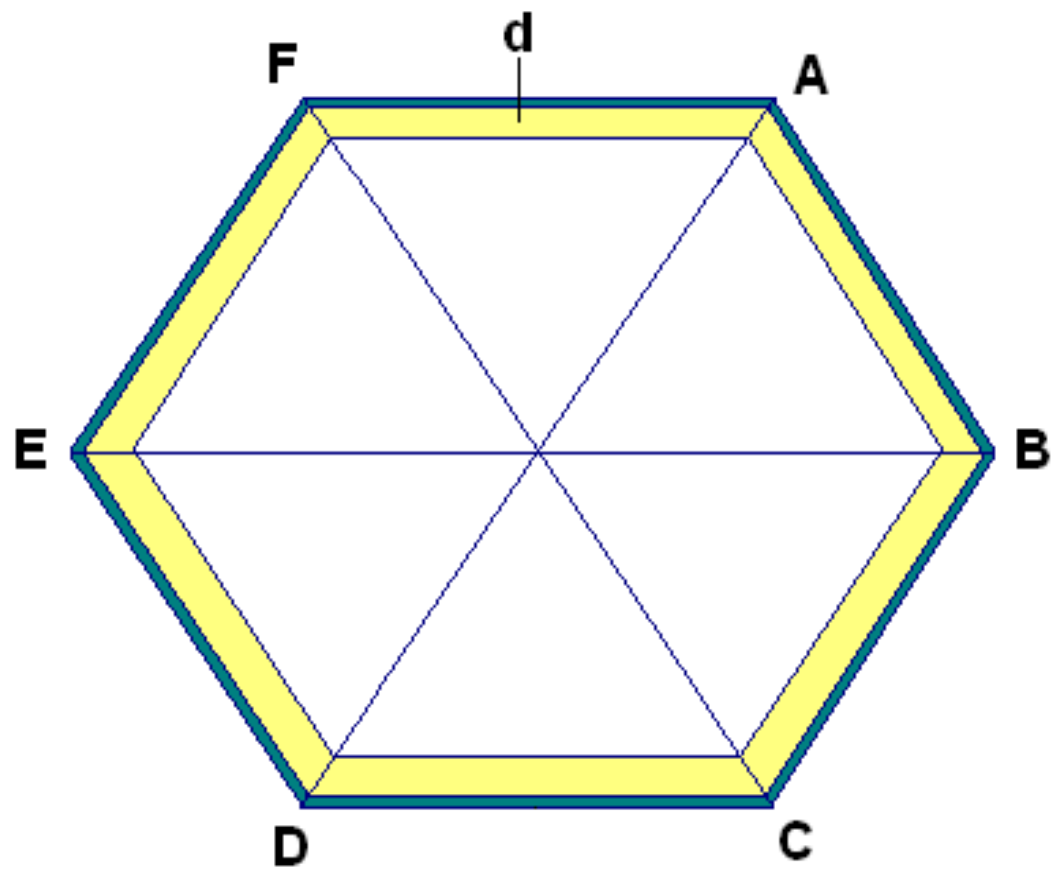

Figure 2

## 3.1 CO-ORDINATE EVALUATION METHOD

The information regarding the latitude and longitude of the MN can be easily got by GPS. After any time interval 'i' let the latitude be $lat_i$ and the longitude $long_i$. Now this latitude and longitude evaluation is done at each interval from the beginning of the initialization phase and after that with a period of 5 ms. The co-ordinate displacement is defined as the change in latitude and longitude after a given time period. It is calculated after any time interval 'i' as the difference in latitudinal co-ordinate equal to $lat_i$-$lat_{i-1}$ and the longitudinal co-ordinate $long_i$-$long_{i-1}$. This are written as $\Delta lat_i$ and $\Delta long_i$ respectively after any time interval't'.

The co-ordinate displacement after each time period is noted down. A new parameter named as the average rate of change of co-ordinate is framed which gives the average of all the co-ordinate displacement corresponding to both the latitude and longitude divided by the total time of observation. These are denoted by the symbols $\lambda_{lat}$ and $\lambda_{long}$ for latitude and longitude respectively and after any time interval 'i' they are given as

$$\lambda_{lat} = ( \sum \Delta lat_i )/ ( 5 * i)$$

$$\lambda_{long} = ( \sum \Delta long_i)/ ( 5 * i)$$

where summation is carried for all integer values starting from 1 to i.

So whenever the MN enters the shaded region as shown in Figure 2, the given algorithm stops. If 'i' number of time intervals have been observed then the co-ordinate of the MN after the time $t_{delay}$ is expected to be given as $((lat_i+ \lambda_{lat}*t_{delay}), (long_i+ \lambda_{long}*t_{delay}))$. By the knowledge of the expected co-ordinate of the MN after the handoff process, it can be expected towards which AP the MN would be heading. So the most potential AP may be scanned, hence the net handoff latency is reduced by a great margin.

To make the approximation more accurate, we calculate the error in the co-ordinate estimation after each time interval. This is carried on after the initialization phase where the expected co-ordinate at the end of the $i^{th}$ interval is compared to the actual co-ordinate after the $i^{th}$ interval. For each latitude and longitude, two errors i.e., the maximum positive error and the maximum negative error are calculated. These are recorded by any sorting algorithm or simply by comparing the error after any time interval with the present values. Let the positive error corresponding to latitude be $pe_{lat}$ and for longitude be $pe_{long}$ and correspondingly for negative error let these be $ne_{lat}$ and $ne_{long}$. These values come into action at the moment of determination of the potential AP. Now the expected range of co-ordinate if 'i' no of time intervals have been observed is given as:

$$(lat_i+ \lambda_{lat}*t_{delay})+ne_{lat} <= \text{Expected latitude} <=(lat_i+ \lambda_{lat}*t_{delay})+ pe_{lat}$$

$$(long_i+ \lambda_{long}*t_{delay})+ne_{long} <= \text{Expected longitude} <=(long_i+ \lambda_{long}*t_{delay})+ pe_{long}$$

So we get a range of values for latitude and longitude. It is shown in the simulation part that it yields at the most 2 APs within this range but gives a higher accuracy corresponding to handover. But most of the time, the algorithm gives one potential AP and in maximum of those cases, the handoff is favorable by scanning the channel corresponding to the most potential AP only. The simulation of algorithm is provided in the next section with a detailed performance evaluation which judges the suitability of this algorithm in real conditions.

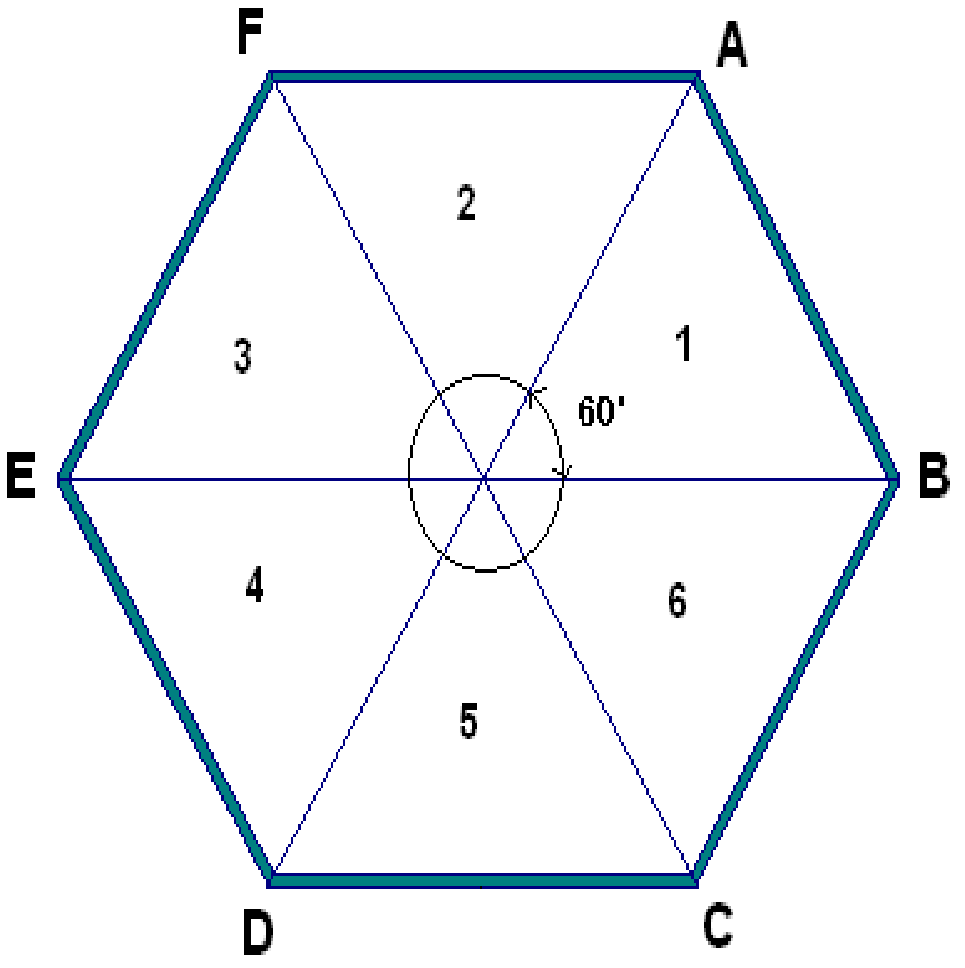

Figure 3

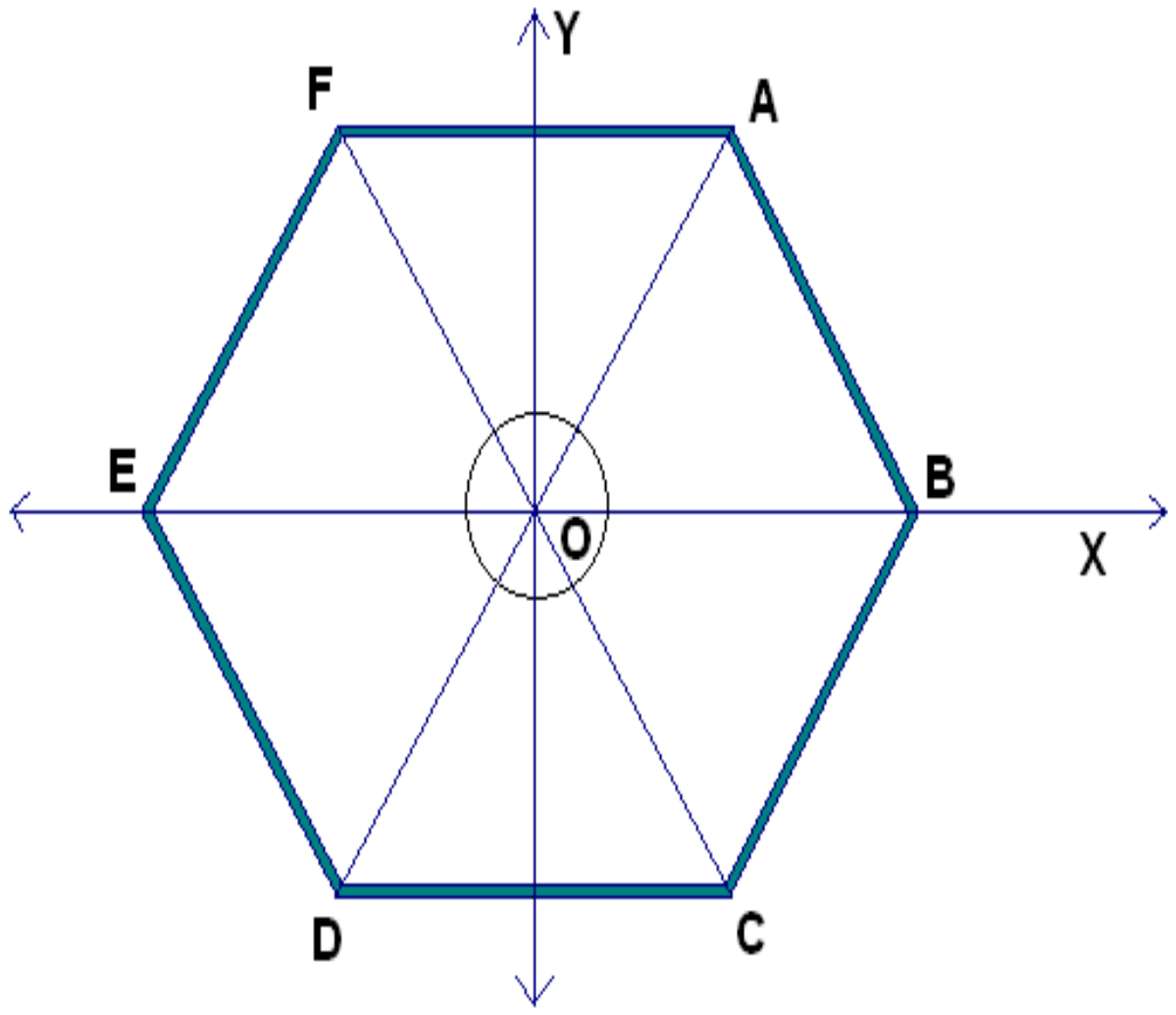

Figure 4

# 4. SIMULATION RESULT

We made a sample run of our algorithm to test the functionality of it. We considered the handover for a MN from the cell in which its call originates. The coverage region of the AP is taken as regular hexagons of length 231m approx (which satisfy the topological conditions of an AP in urban area). At the end of the algorithm we note the range of co-ordinate in which the MN may lie after handover and we compared this with its actual co-ordinate that it makes with the previous AP. The result of this simulation justifies the appropriateness of our algorithm.

The average scan phase was taken as 50 ms. All the co-ordinates are in meters and are measured in reference to the present AP as the origin and axis shown as in Figure 4. The average speed of the MN was recorded as 19.0222 m/s after the origination of call. As proposed when it reaches the boundary, as shaded in Figure 2 the algorithm stops. In this sample run incorporated with our algorithm we get the range of expected co-ordinate in which the MN may lie after our algorithm is given as 199.9185 to 199.9070 for x co-ordinate and 77.4862 to 77.4786 for y co-ordinate (co-ordinates measured according to the system of axis as in Figure 4). This indicates that it is heading towards AP 1 (refer to Figure 4 by considering the edge length of the hexagon as 231 m) as the expected range of co-ordinates lie in the coverage area of AP 1. So only AP1's channel may be scanned during handover which does reduces the handover latency to a great extent. After the handover phase the recorded actual co-ordinate of the MN is (199.9117, 77.4825). This signifies the appropriate- ness of our algorithm. A set of sample runs were made using this algorithm by varying the parameters like cell coverage area or the speed of the MN which yielded the result as 99.25% of cases in which the correct AP or AP's were identified. In approximately 22.50% of cases there were two AP's to be scanned and for those cases the scanning phase was taken double of the scanning phase of each AP. A graph showing the error in recording the co-ordinate of the MN after each time interval is shown in Figure 5 and 6 for x and y co-ordinate respectively. For this run we compared the actual co-ordinate of the MN with the expected co-ordinate for a specific 25ms time period covering 5 iterations in Figure 7 and Figure 8 for x and y co-ordinate respectively. (Green represent the expected and red the actual angle). The varying speed of the MN and the minimum distance'd' before which the algorithm must be stopped is shown in Figure 9 and Figure 10 respectively.

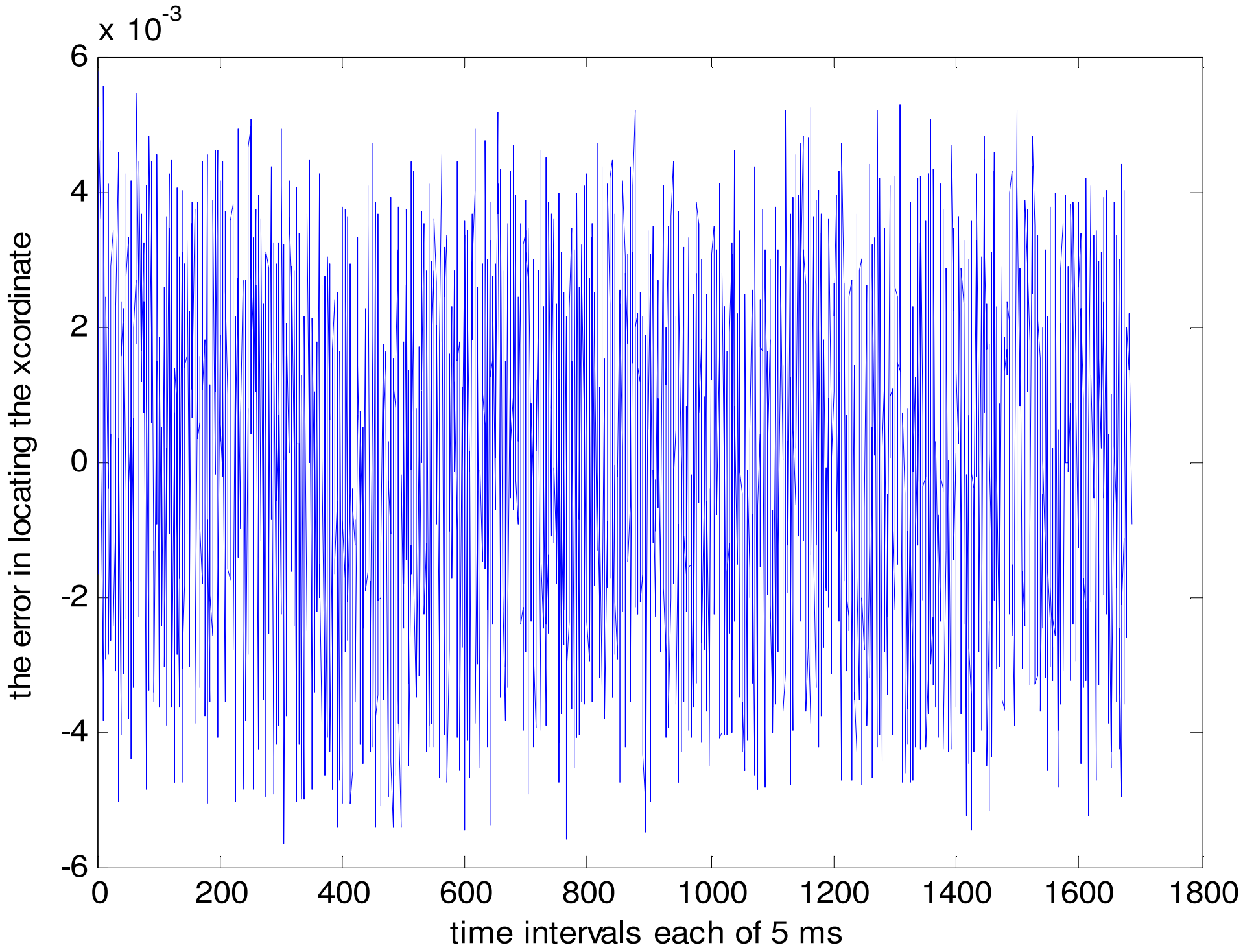

Figure 5.Plot of the error in locating x co-ordinate (in meter) vs the time intervals (each interval of 5 ms)

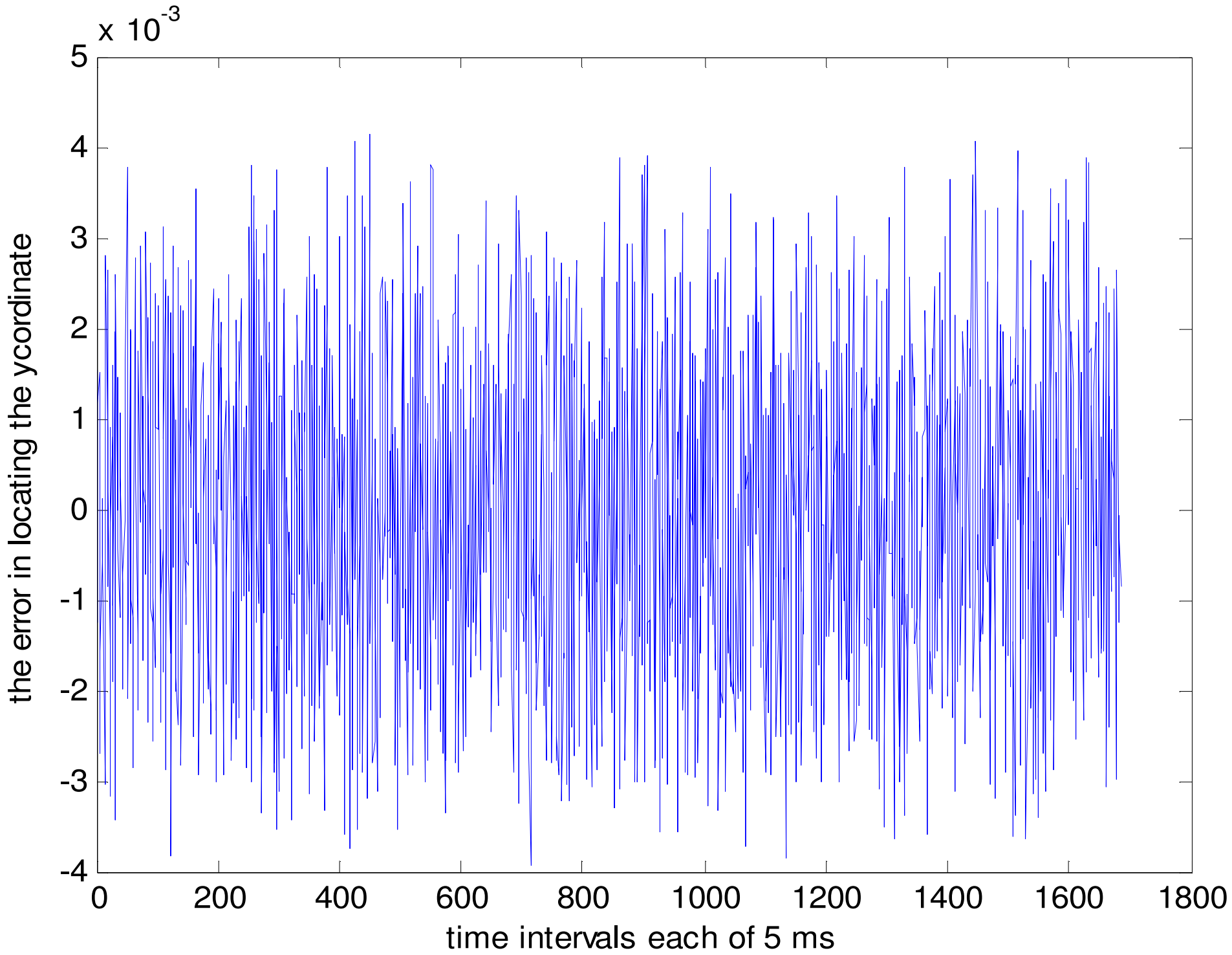

Figure 6.Plot of the error in locating y co-ordinate (in meter) vs the time intervals (each interval of 5 ms)

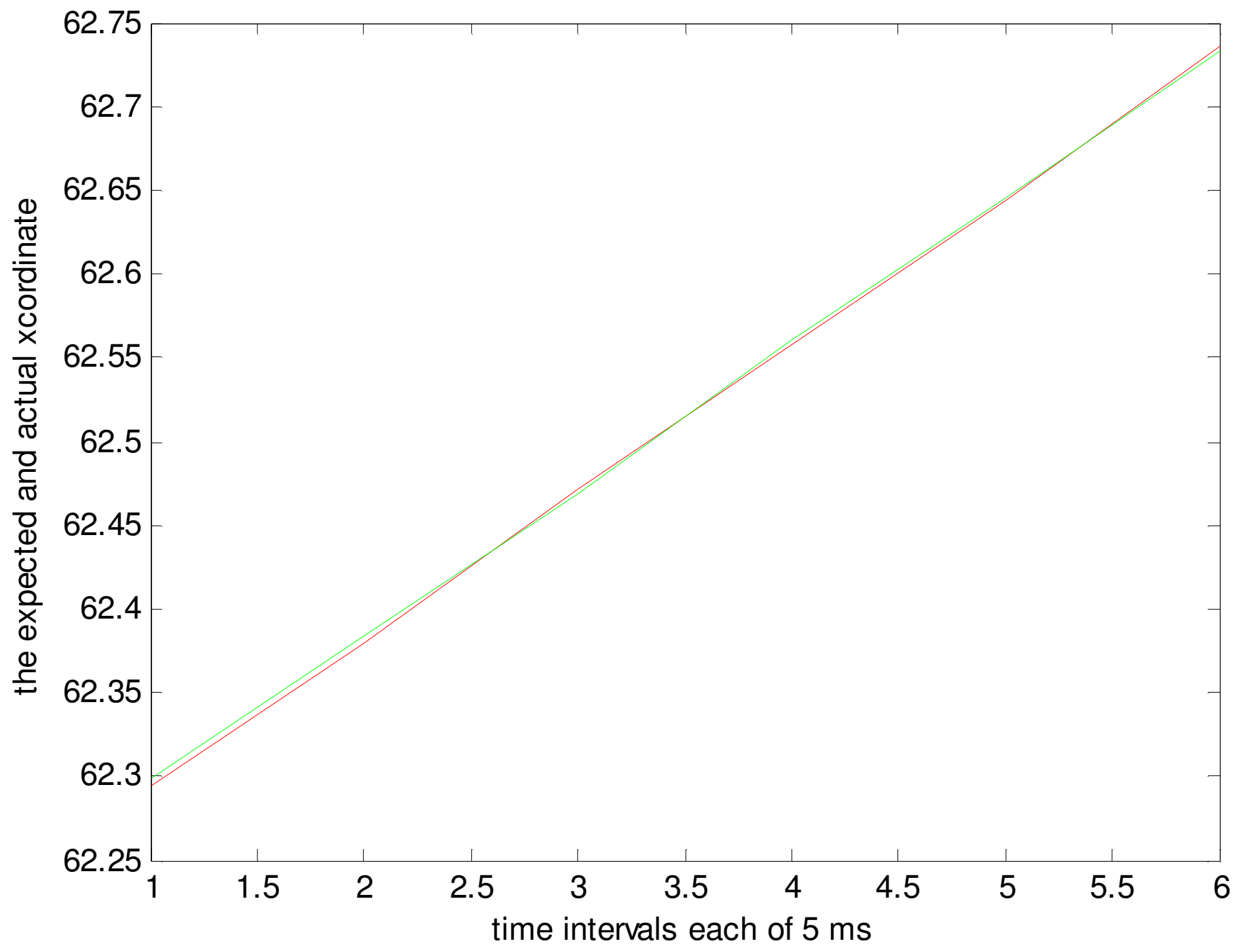

Figure 7.A magnified region of any 5 time interval in which the expected and the actual x co-ordinate (in meters) are plotted.

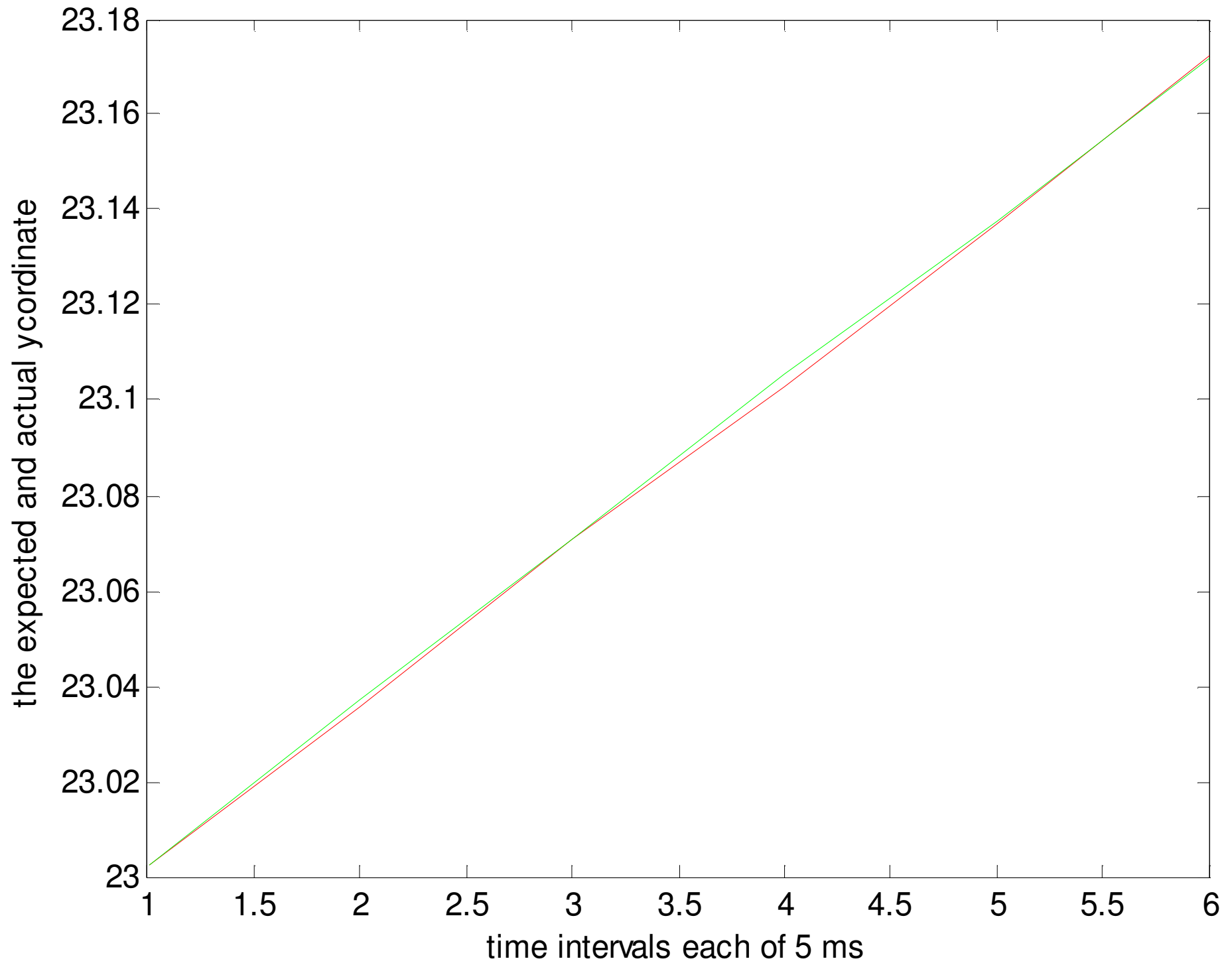

Figure 8.A magnified region of any 5 time interval in which the expected and the actual y co-ordinate (in meters) are plotted

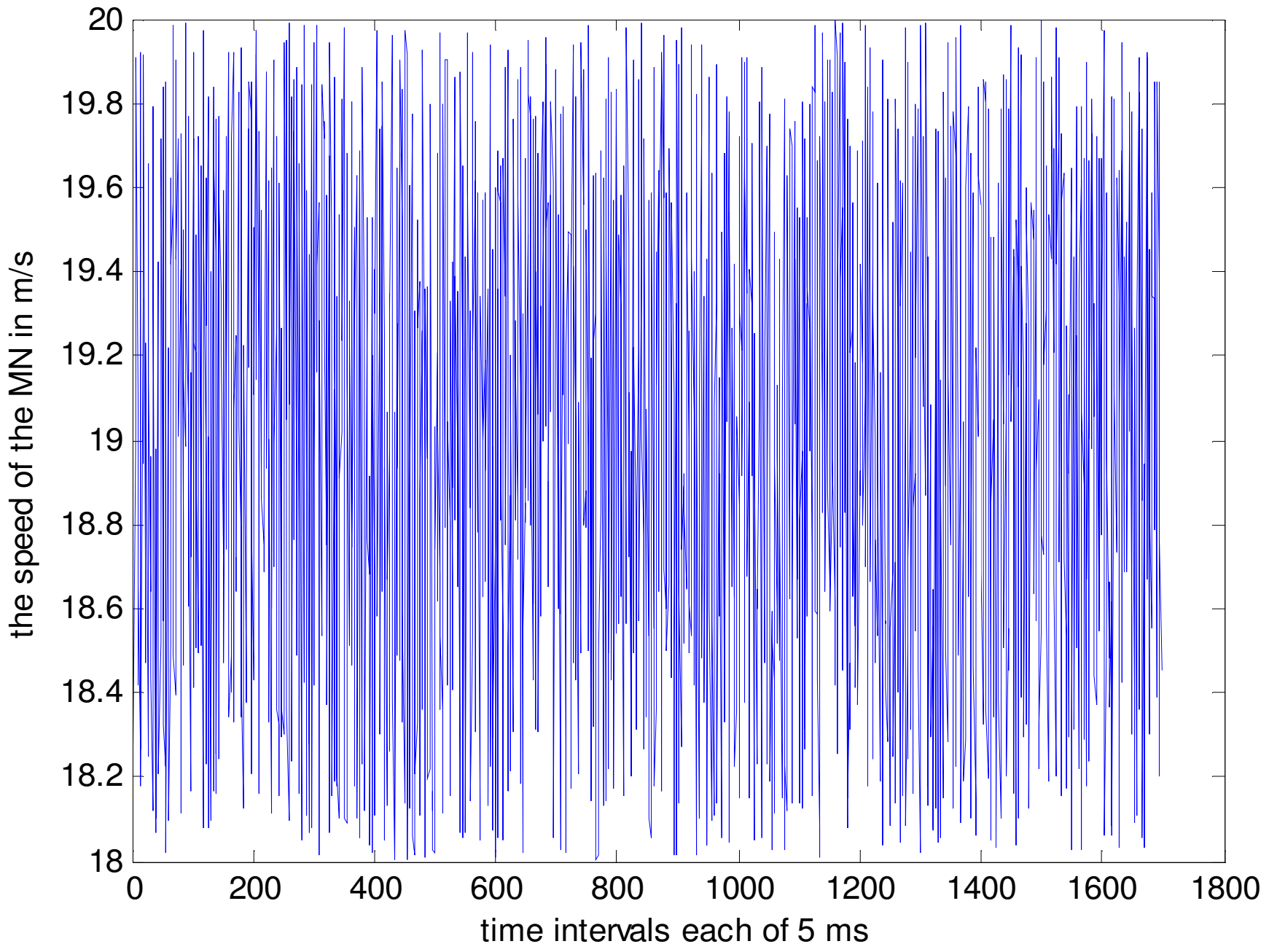

Figure 9.Plot of speed of the MN vs the time interval

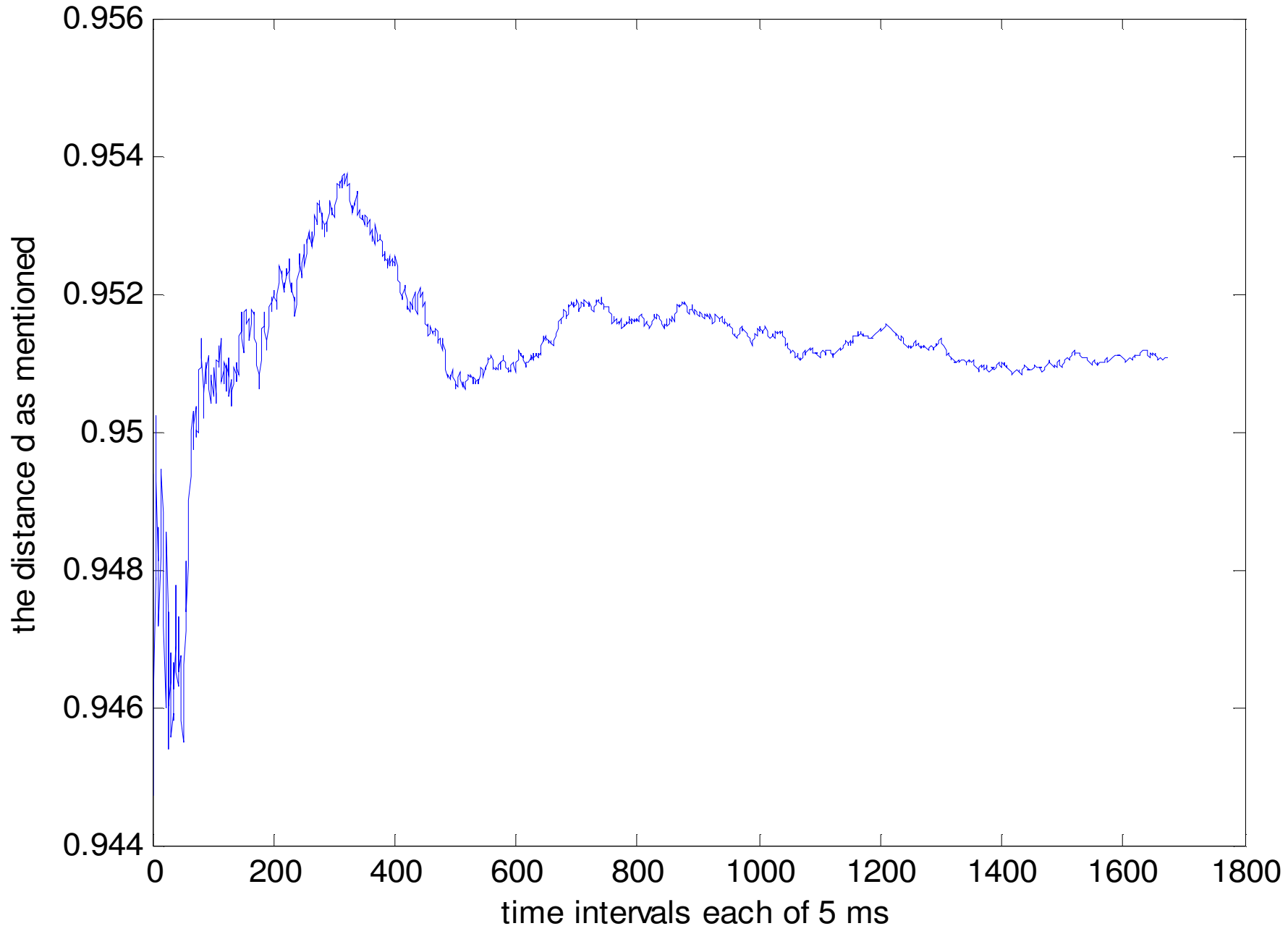

Figure 10.Plot of distance ‘ d’ vs the time interval

When a MN moves out of its old AP’s coverage area it can easily find out the most potential AP by our proposed algorithm. If the handoff process is initiated by comparing the relative signal strength of the old AP and its neighboring APs there remains a chance of false handoff. Let at the boundary region, the relative signal strength of AP2 is greater than that of AP1, but the direction of the MN’s movement is actually towards AP1. In this case to take the right decision for handoff initiation is quite difficult. This problem can be easily handled by our algorithm as it takes care of the change in direction of velocity of the MN.

# 4. CONCLUSION AND FUTURE WORKS

By this co-ordinate evaluation method, we can reduce the handoff latency to a great deal as we can get a clear idea as to which channel to scan for a particular MN. The selection of the most potential AP by the MN effectively reduces the scanning delay, as the number of channels scanned will be lower. However, this method may prove erroneous if the MN follows a haphazard trajectory in which we cannot get a clear estimate of its future position even if we know the present trajectory using GPS.

Future work regarding this topic may include researches aiming to minimize the error approximation to a greater extent. We have tried for accurate approximation as far as possible, but there is always scope for improvement.

# 5. REFERENCES

[1] W. Puangkor and P. Pongpaibool. 'A Survey of Techniques for Reducing Handover Latency and Packet Loss in Mobile IPv6'. IM 20060306.

[2] H-Soo Kim et. al. 'Selective Channel Scanning for Fast Handoff in Wireless-LAN Using Neighbor-graph' Japan, July2004. International Technical Conference on Circuits/Systems Computers and Communication.

[3] C-Sheng Li et.al. 'A Neighbor Caching Mechanism for Handoff in IEEE 802.11 Wireless networks' Springer 20 March 2008,DOI 10.1007/s11227-008-0175-3.

[4] A. Dutta, S Madhani, Wai Chen, "GPS-IP Based Fast Handoff for Mobiles".

[5] C-Chao Tseng, K-H Chi, M-D Hsieh and H-H Chang," Location-based Fast Handoff for 802.11 Networks",IEEE COMMUNICATIONS LETTERS,VOL9, NO 4 April 2005.

[6] S. Kyriazakos, D. Drakoulis and G. Karetsos, "Optimazation of The Handover Algorithm Based on The Position of The Mobile Terminals" in Proceedings of Symposium on Communications and Vehicular Technology, October 2000.

[7] J. Pesola and S. Pokanen, "Location-aided Handover in Heterogeneous Wireless Networks," in Proceedings of Mobile Location Workshop, May 2003.

[8] P-Jung Huang and Y-Chee Tseng. 'A Fast Handoff Mechanism for IEEE 802.11 and IAPP Networks'. Vehicular Technology Conference,2006.VTC 2006-Spring.IEEE 63rd,vol 2,pp 966-970.

[9] J. Montavont and T. Noel, " IEEE 802.11 Handovers Assisted by GPS Information", IEEE 1-4244-0495 9/06,2006.

[10] S. Mohanty and I. F.Akylidiz "A Cross Layer (Layer 2+3) Handoff Management Protocol for Next-Generation Wireless Systems", IEEE TRANSACTIONS ON MOBILE COMPUTING,Vol-5,No-10 OCT 2006.

## AUTHOR BIOGRAPHIES

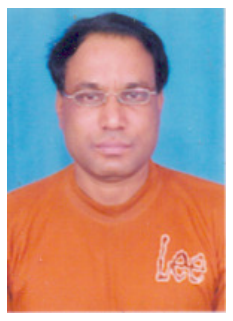

**Debabrata Sarddar** is currently pursuing his PhD at Jadavpur University. He completed his M.Tech in Computer Science & Engineering from DAVV, Indore in 2006, and his B.E in Computer Science & Engineering from NIT, Durgapur in 2001. He was earlier a lecturer at Kalyani University. His research interest includes wireless and mobile system.

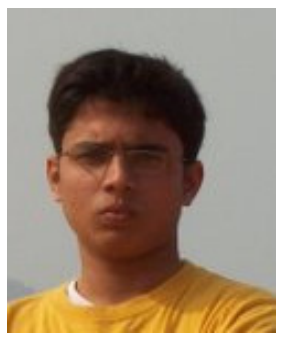

**Joydeep Banerjee** is presently pursuing B.Tech Degree in Electronics and Telecommunication Engg. at Jadavpur University. His research interest includes wireless sensor networks and wireless communication systems.

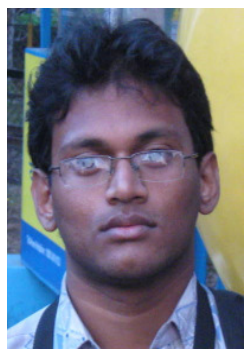

**Souvik Kumar Saha** is presently pursuing B.Tech Degree in Electronics and Telecommunication Engg. at Jadavpur University. His research interest includes wireless sensor networks and wireless communication systems.

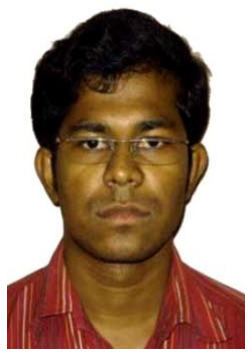

**Tapas Jana** is presently pursuing B.Tech Degree in Electronics and Communication Engg. at Netaji Subhash Engg. College, under West Bengal University Technology. His research interest includes wireless sensor networks and wireless communication systems

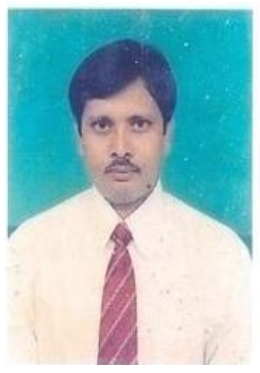

**Utpal Biswas** received his B.E, M.E and PhD degrees in Computer Science and Engineering from Jadavpur University, India in 1993, 2001 and 2008 respectively. He served as a faculty member in NIT, Durgapur, India in the department of Computer Science and Engineering from 1994 to 2001. Currently, he is working as an associate professor in the department of Computer Science and Engineering, University of Kalyani, West Bengal, India. He is a co-author of about 35 research articles in different journals, book chapters and conferences. His research interests include optical communication, ad-hoc and mobile communication, semantic web services, E-governance etc.

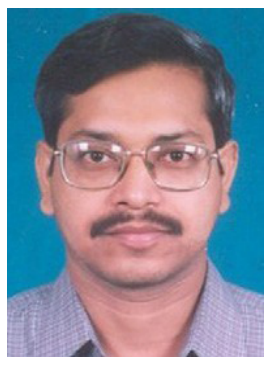

**Mrinal Kanti Naskar** received his B.Tech. (Hons) and M.Tech degrees from E&ECE Department, IIT Kharagpur, India in 1987 and 1989 respectively and Ph.D. from Jadavpur University, India in 2006.. He served as a faculty member in NIT, Jamshedpur and NIT, Durgapur during 1991-1996 and 1996-1999 respectively. Currently, he is a professor in the Department of Electronics and Tele-Communication Engineering, Jadavpur University, Kolkata, India where he is in charge of the Advanced Digital and Embedded Systems Lab. His research interests include ad-hoc networks, optical networks, wireless sensor networks, wireless and mobile networks and embedded systems.

He is an author/co-author of the several published/accepted articles in WDM optical networking field that include "Adaptive Dynamic Wavelength Routing for WDM Optical Networks" [WOCN,2006], "A Heuristic Solution to SADM minimization for Static Traffic Grooming in WDM uni

-directional Ring Networks" [Photonic Network Communication, 2006],"Genetic Evolutionary Approach for Static Traffic Grooming to SONET over WDM Optical Networks" [Computer Communication, Elsevier, 2007], and "Genetic Evolutionary Algorithm for Optimal Allocation of Wavelength Converters in WDM Optical Networks" [Photonic Network Communications,2008].